\documentclass[aps,prl,amsfonts,amssymb,twocolumn,epsfig,graphics,showpacs]{revtex4}

\usepackage{graphicx}
\usepackage{epsfig}

\newcommand \be {\begin{equation}} 
\newcommand \ee {\end{equation}}

\newcommand \bea {\begin{eqnarray}} 
\newcommand \eea {\end{eqnarray}}

\begin{document}

\newcommand{\<}{\langle}
\renewcommand{\>}{\rangle}
\newcommand{\ua}{\uparrow}
\newcommand{\da}{\downarrow}
\newcommand{\cs}{$\clubsuit$}

\def\br{\mbox{\boldmath $r$}}

\date{\today}

\title{The Phase Diagram of Random Heteropolymers}
\author{ A. Montanari$^{\,1}$, M.\ M\"uller$^{\,2}$, M. M\'ezard$^{\,2}$}
\affiliation{
$^{1\,}$Laboratoire de Physique Th\'{e}orique de l'Ecole Normale
  Sup\'{e}rieure\footnote {UMR 8549, Unit{\'e}   Mixte de Recherche du 
  Centre National de la Recherche Scientifique et de 
  l' Ecole Normale Sup{\'e}rieure. }
}
\affiliation{
$^{2\,}$Laboratoire de Physique Th\'eorique et Mod\`eles Statistiques,\\
Universit\'e Paris-Sud, b\^atiment 100, F-91405 Orsay,
France.
}


\begin{abstract}
We propose a new analytic approach to study the phase diagram of random
heteropolymers, based on the cavity method. For copolymers we analyze the
nature and phenomenology of the glass transition as a function of sequence
correlations. Depending on these correlations, we find that two different
scenarios for the glass transition can occur. 
We show that, beside the much studied possibility
of an abrupt freezing transition at low temperature, the system
can exhibit, upon cooling, a first transition to a soft glass phase with fully
broken replica symmetry and a continuously growing degree of freezing as the
temperature is lowered. 
 
\end{abstract}

\maketitle

The statistical mechanics of random heteropolymers has attracted a lot of 
interest in the past two decades, mainly for its close relation to the long 
standing problem of protein folding~\cite{Grosberg00}, 
but also because of its applications to material 
sciences~\cite{SfatosShakhnovich97}. It has soon been recognized that 
the energy landscape and low temperature physics of heteropolymers bear 
a strong resemblance with that of spin glasses, and many theoretical concepts 
have been fruitfully adapted to that particular class of disordered 
systems~\cite{WolynesREM, MezardParisi87b}.

A popular guiding idea in this field was the conjecture that
typical low energy configurations of a random heteropolymer behave as
independent energy levels \cite{PolymerREM} as  it
happens in the random energy model (REM) \cite{Derrida81}. Mean field replica
calculations corroborate such a discontinuous glass
transition with one step of replica symmetry breaking, both for random bond
interactions \cite{ShakhnovichGutin89GarelOrland88} and for copolymers
\cite{GarelLeibler94SfatosGutin93}. However, the applicability of the simple REM-like picture to heteropolymers
and proteins has been questioned \cite{GrosbergREM}.
The numerical study of self-consistent dynamical equations
for short chains~\cite{TimoshenkoDawson} hints at a more refined scenario which
we shall compare with our own results below.

The polymer freezing manifests itself in persisting correlations between two times 
$t_w$ and $t_w+t$, after the molecule has been 
prepared in a random conformation at time 
$0$ ~\cite{dyn_rev}. The following family of correlation 
functions for a chain of length $N$ proves very 
useful~\cite{CopolymerIsing}: 
$C_l(t,t_w) = N^{-1}\sum_i 
\< \Delta \br_{i,l}(t)\cdot \Delta\br_{i,l}(t_w)\>$ . Here
$\Delta \br_{i,l} = \br_{i+l}-\br_i$ and $\br_i$ is the spatial position of
the $i$-th monomer of the chain. 
Within the glassy phase, these correlation functions 
should exhibit a {\it plateau} at $q_l=\lim_{t\to\infty}\lim_{t_w\to\infty}
C_l(t,t_w)$ ($t$ and $t_w$ are much smaller than 
the equilibration time $\tau_{\rm eq}$).

The REM-like scenario predicts an abrupt freezing in a metastable conformation, implying a discontinuous jump
of the plateau height $q_l$ from zero to a finite value at the glass transition.
However, many other disordered systems, like spin
glasses, exhibit a continuous glass transition (in the "overlap" parameter, $q_l$). This has recently been predicted for the random amphiphilic chain~\cite{CopolymerIsing}, as does our analysis for a certain class of correlated sequences. The folding properties of a given heteropolymer will be crucially affected by the type of glass transition it undergoes, since the
dynamics (e.g., the aging behavior) are rather different in these
two scenarios~\cite{dyn_rev}. 

Apart from numerical simulations, most of our fundamental knowledge about heteropolymers relies on two main analytical methods: the replica approach, with some kind of Gaussian variational Ansatz~\cite{ShakhnovichGutin89GarelOrland88}, and the use of dynamical equations, with some self-consistent closure approximation~\cite{Copolymerdyn,TimoshenkoDawson}. Considering that these methods use some rather severe approximations which are not really independent from each other, it is clearly useful to have a new analytic method emerging.
We propose here such a new approach, based on the cavity method \cite{cavityT}, which is an extension of the classical Bethe approximation
to frustrated systems. It is the simplest
mean-field approximation that handles exactly the
short distance correlations~\cite{Bethehomopolymer}, and 
in this respect it 
is very different from the usual replica approach. 
Moreover, as we will see below, the existence of a non-trivial
statistical mechanics model on a 'Bethe lattice' for which the method is exact is very helpful.
We expect the cavity approximation to be faithful as long as 
crystalline, or more complex, spatially ordered structures are 
suppressed by the randomness of the chain sequence.

\begin{figure}
	\resizebox{8.5 cm}{!}{
  \includegraphics{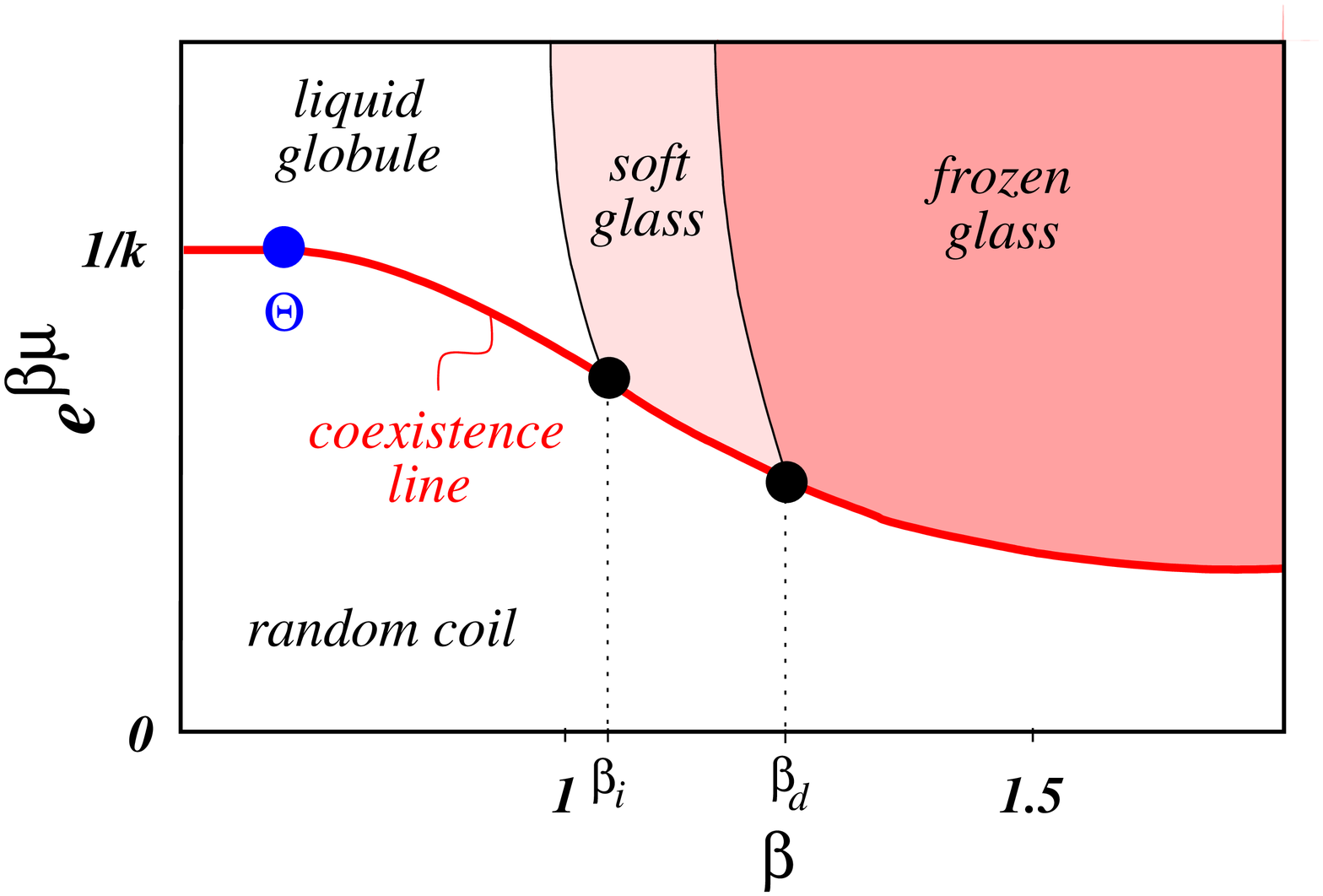}}
	\resizebox{8.5 cm}{!}{
  \includegraphics{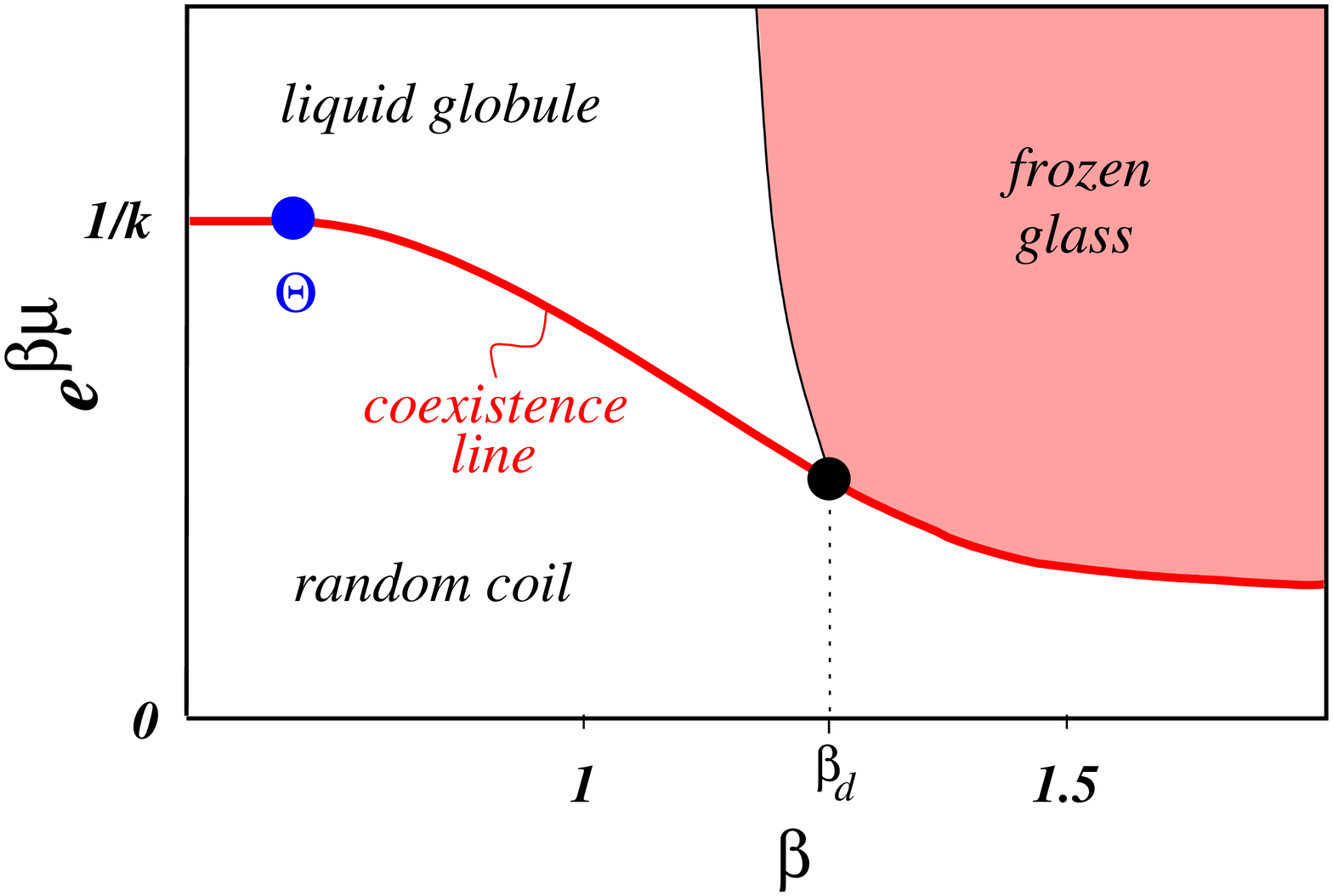}}	
\caption{The phase diagram in the grand canonical ensemble. 
An unconstrained polymer lives on the coexistence line
($\mu=\mu_c(T)$). 
In correlated
 HP-polymers and anticorrelated ampholytes (top) a continuous glass transition 
occurs at the liquid instability. It is followed 
by a discontinuous freezing transition. Oppositely correlated sequences
(bottom) exhibit a direct freezing transition from the liquid globule.}
\label{phasedia}
\end{figure}

\begin{figure}
	\resizebox{8.5 cm}{!}{
  \includegraphics{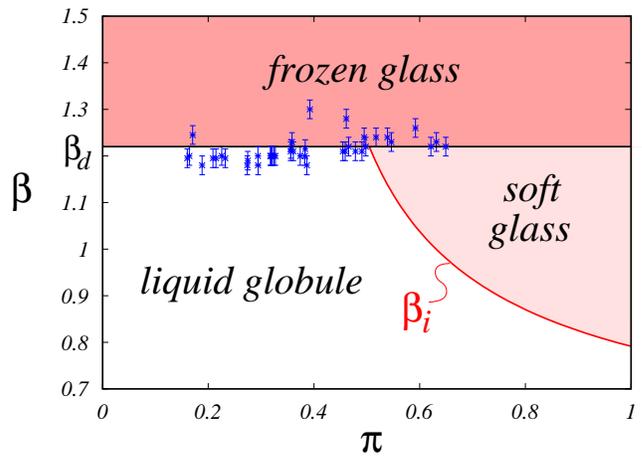}}
\caption{The phase diagram as a function of sequence correlations $\pi$ 
for  HP polymers. At high temperature (small $\beta$) the polymer 
is found in a liquid globule phase. For correlated sequences ($\pi\gtrsim 0.5$),
the molecule undergoes a continuous phase transition at
inverse temperature $\beta_i$ given by (\ref{ABinstab}), followed by a
discontinuous freezing at $\beta_d$. Anticorrelated sequences ($\pi\lesssim
0.5$) freeze discontinuously at $\beta_d$ without intermediate phase. The
inverse temperature of abrupt freezing, $\beta_d$, turns out to be nearly independent of the sequence, as indicated by the horizontal line at  $\beta_d\approx 1.23$. The data points show the $\beta_d$'s computed for a few particular 
sequences with $L=20$.}
\label{transitions}
\end{figure}

We show that for fixed monomer-monomer interactions,
both types of glass transition can be found.
Depending on the correlations in the chain sequence, one finds
either a transition to a `frozen glass' phase, with a jump in $q_l$
(we shall call $T_d=1/\beta_d$ the corresponding transition temperature), or a
continuous transition to a 'soft glass' phase, where $q_l$ develops
continuously from zero (we denote by $\beta_i$ the inverse critical 
temperature in this case), see Figs.~\ref{phasedia} and \ref{transitions}.
 The predicted soft glass phase
bears some resemblance with the intermediate glass phase found in the dynamic
approaches~\cite{TimoshenkoDawson}.

Consider a lattice polymer whose conformation is given by a self-avoiding 
walk. Two monomers interact with an energy $e_{\sigma,\sigma'}$ 
if they occupy two neighboring sites $i$ and $j$. Here $\sigma,\sigma'\in
\{1,-1\}\leftrightarrow \{A,B\}$ denote the monomer types. We will restrict to
the symmetric case $ e_{AA}= e_{BB}=- e_{AB}$ and only
consider neutral polymers (equal number of $A$'s and $B$'s). The case
$ e_{AA}=-1$ is close to the popular HP-model for proteins where
hydrophobic and polar constituents tend to cluster, while $ e_{AA}=1$
corresponds to charged polymers (ampholytes) with short-range interactions.

On a lattice with $V$ sites, we consider the polymer in the grand canonical
ensemble with a chemical potential $\mu$ per monomer. One can identify two
phases depending on the scaling of the average length $\<N\>$ of the
polymer \cite{Bethehomopolymer}. For small $\mu$ ($\mu<\mu_c(T)$) 
there exists an
infinitely diluted phase where $\<N\>/V \to 0$ as $V \to
\infty$. For $\mu>\mu_c$, there is a dense phase with a finite $\<N\>/V$. To
describe a 
polymer in equilibrium with the solvent the chemical
potential has to be adjusted to the critical value $\mu_c$. 
The $\Theta$-point where the 
polymer collapses to the liquid
globule occurs as a tricritical point on this line \cite{DeGennes75}, 
see Fig.~\ref{phasedia}.

%
Before explaining the cavity approach, we notice 
that its approximations 
become exact on a `Bethe lattice' 
whose appropriate definition for a frustrated system is a random graph with $N$ vertices and fixed 
connectivity (taken equal to $k+1\equiv 6$ here to mimic 
cubic lattices) \cite{cavityT}. Since
the size of typical loops in such a graph diverges as $O(\log(N))$, the graph
is locally tree-like, allowing for iterative solutions of the statistical
physics problem.

For technical reasons we work with a periodic sequence of monomers with period
$L$ and take the limit $L\rightarrow \infty$ in the end. Let us now consider an
oriented edge of the graph, going from a `root' site R to another site S. It
can be in one of the following states: $(0)$: neither root nor edge are
occupied; $(+,a)$ or $(-,a)$: the root is occupied by monomer $a\in [1,L]$
and the backbone continues on the edge, the sign $\pm$
indicating that S is occupied by monomer $a\pm 1$;
 $(2a)$: the root is occupied by
monomer $a$, but the edge is free. We now exploit the local tree-like
structure to recursively express the (Gibbs) probability $p_{\alpha}^{(0)}$ of
finding the oriented edge `0' in the conformation $\alpha \in \{0,(+,a),(-,a), 2a\}$ in terms of the corresponding probabilities $p_{\alpha'}^{(i)}$
on the remaining edges $(i=1,\dots,k)$ that are connected to
 the root of the edge `0':
\bea
	\label{cavity0}
	p_0^{(0)} &=&C^{-1}\prod_{i=1}^{k} (p_0^{(i)}+\sum_{a'=1}^L p_{2a'}^{(i)}),\\
	p_{\pm, a}^{(0)} &=&C^{-1}e^{\beta \mu}\sum_{i=1}^{k} p_{\pm, a\mp 1}^{(i)} \prod_{j\ne i}^{k} \psi_a^{(j)},\\
	\label{cavity2}
	p_{2a}^{(0)} &=&C^{-1}e^{\beta \mu}\sum_{i_1\ne i_2}^{k} p_{+,a-1}^{(i_1)} p_{-,a+1}^{(i_2)} \prod_{j\ne i_1,i_2}^{k} \psi_a^{(j)},
\eea
where $C$ is a normalization and we have introduced
$\psi_a^{(j)}=p_0^{(i)}+\sum_{a'=1}^L p_{2a'}^{(j)}e^{-\beta e_{\sigma_a,\sigma_{a'}}}$.

The recursion is exact on a tree, but it also holds asymptotically on a large
$N$ Bethe lattice, in the liquid phase or inside a pure state 
where the probabilities on the $k$ parent links are uncorrelated.

The liquid phase is described by a fixed point 
$p_\alpha^{(i)}\equiv p_\alpha^{*}$ of Eqs.~(\ref{cavity0}-\ref{cavity2}). 
The liquid free energy $\phi_{\rm liq}$ is obtained as a sum of site and 
edge contributions which are expressed in terms of $\{ p^*_{\alpha}\}$
following the general prescription of Ref. \cite{cavityT}.
It turns out that the liquid globule of neutral copolymers bears no 
trace of disorder since $\phi_{\rm liq}$ is independent of the sequence.

For any heterogeneous sequence the entropy of the liquid solution turns 
negative at sufficiently low temperatures, indicating the existence of a 
glass phase which breaks the translational invariance. The latter is also 
clearly observed in Monte Carlo simulations both on the Bethe and the cubic 
3d lattice \cite{longpaper}. To treat the glass phase, the above framework 
has to be extended. We assume that below the glass transition the phase 
space splits into a large number of pure states. Restricting the Gibbs 
measure to a state $\gamma$ gives rise to probabilities 
$p_{\alpha}^{(i;\gamma)}$, and we again obtain a closed recursion 
relation if we look at the distribution of local probabilities over all 
pure states \cite{cavityT}, $\rho({\bf p})={\mathcal{N}}^{-1}\sum_{\gamma}
w_\gamma\delta({\bf p}-{\bf p}^{(i;\gamma)})$.
It will not depend on the site $i$ since the graph is regular. The cavity 
recursion then reads
\bea
	\label{onestepcavity}
	\rho({\bf p})=\frac{1}{\mathcal{Z}}\int \prod_{i=1}^k \!\!\!\!&&\rho({\bf p}^{(i)})d{\bf p}^{(i)} \,\,\,\delta({\bf p}-{\bf p}^{(0)}[{\bf p}^{(1)},\dots,{\bf p}^{(k)}])\nonumber\\
	&&e^{-m\beta\Delta f[{\bf p}^{(1)},\dots,{\bf p}^{(k)}]}
\eea
where ${\bf p}^{(0)}[{\bf p}^{(1)},\dots,{\bf p}^{(k)}]$ is the pure 
state recursion (\ref{cavity0}-\ref{cavity2}). 

We have introduced in (\ref{onestepcavity}) a 
reweighting with respect to the free energy change $\Delta f$ corresponding 
to the addition of the site $0$, in order to take into account the exponential 
increase of the number of pure states as a function of the total free energy. 
$\Delta f$ is related to the normalization factor in 
(\ref{cavity0}-\ref{cavity2}) 
by $\exp(-\beta\Delta f)=C$. The reweighting depends on
$m\in [0,1]$ which is the 
cavity analog 
of the breakpoint
parameter appearing in Parisi's one-step replica symmetry 
breaking scheme. As explained in \cite{cavityT,longpaper}, one can 
write down an $m$-dependent free energy  $\phi_1(m)$, and
the dominant metastable states correspond to $m^*$ which maximizes $\phi_1(m)$.

The local stability of the liquid phase with respect to the glass transition
can be studied by computing appropriately defined glass susceptibilities.
The simplest one can be written  by associating to each
site $i$ of the lattice a `spin' $s_i$. We define $s_i=0$ if $i$ is empty
and $s_i = \sigma$, if $i$ is occupied by a monomer of type $\sigma$.
The  susceptibility is given by 
$\chi_g = V^{-1}\sum_{ij}\overline{\<s_i s_j\>^2}$.  A divergence 
of $\chi_g$ signals the appearance of a glass phase.

The instability of the liquid can be analyzed for arbitrary 
sequences of monomers. It occurs at the temperature $\beta_i$ determined by
\be
	\label{ABinstab}
	\frac{1+z_2(\beta_i) \cosh(\beta_i)}{z_2(\beta_i) \sinh(\beta_i)}=
	-k^{1/2}+\frac{2(k-1)Q_{\rm seq}}{k^{1/2}(1-k^{-L/2})},
\ee
where $z_2(\beta_i)=Lp_2^*/p_0^*$ is independent of the 
sequence~\cite{longpaper}. The latter only enters through the term 
$Q_{\rm seq}=\sum_{b=0}^{L-1}q_b(\pm \sqrt{k})^{-b}$ where 
$q_b=1/L\sum_{a=1}^L\sigma_a\sigma_{a+b}$ is the sequence autocorrelation 
function and the upper/lower sign corresponds to the HP-model and the 
ampholyte, respectively. Equation (\ref{ABinstab}) can be regarded as 
the generalization of a classic result by de Gennes \cite{MPSdeGennes}
to strongly frustrated systems. 

In Fig.~\ref{transitions} we plot the local instability $\beta_i$ 
obtained from Eq. (\ref{ABinstab})
as a function of sequence correlations in the simple case of Markovian 
chains described by the probability $\pi$ for two consecutive monomers to 
be equal. If the instability occurs too close to the entropy crisis of the 
liquid, it will be prevented by a discontinuous transition at $\beta_d$ 
where a
non-trivial solution of the cavity equation (\ref{onestepcavity}) appears.
This effect can be studied  by a numerical solution of this equation
using a population
dynamics method \cite{cavityT,longpaper}. Upon increasing the period $L$, 
we found the transition points to approach a correlation-independent 
constant $\beta_d\approx 1.23\pm 0.03$ that we believe to be a good 
approximation in the large $L$ limit, see Fig. \ref{transitions}. 
 
For ampholytic sequences 
with a bias to alternation, $\pi\lesssim 0.5$, and for HP-like polymers with a 
preference for blocks, $\pi\gtrsim 0.5$, 
there is a continuous glass transition 
at $\beta_i < \beta_d$ as predicted by (\ref{ABinstab}). It is 
followed by a discontinuous freezing transition at lower temperature 
$\beta_d$.  In copolymers with opposite 
correlations, the liquid instability is irrelevant and the freezing transition 
occurs directly from the liquid globule phase. The grand-canonical
phase diagrams for the two scenarios are presented in Fig.~\ref{phasedia} for two fixed values of $\pi$.  

At the continuous phase transition $\beta_i$, the distribution $\rho({\bf p})$ 
continuously departs from the liquid fixed point $\delta({\bf p}-{\bf p}^*)$. 
The transition is of third order in the free energy, and 
as usual, the presence of a  continuous transition indicates that the replica 
symmetry is fully broken. The states will organize in an ultrametric 
hierarchy of clusters with small `preferences' towards some particular
conformations.
On lowering the temperature, these preferences grow 
and result in increasing correlations at larger distances. 
This effect might be related to the scale-dependent freezing predicted in 
\cite{Copolymerdyn,CopolymerIsing} and the intermediate 
glass phase found in the numerics of \cite{TimoshenkoDawson}.

The frozen phase at low temperature is of rather different character. 
The site probabilities $p^{(i)}_{\alpha}$ are strongly biased towards 
one particular site-dependent conformation. The pure states correspond to 
almost frozen conformations and have very small internal entropy, as in the 
REM. 

\begin{figure}
	\resizebox{8.5 cm}{!}{
  \includegraphics{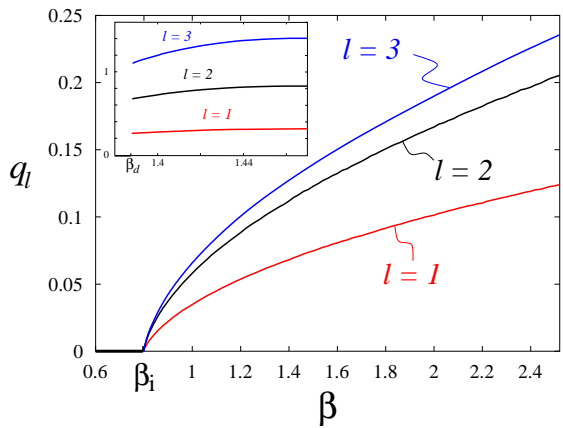}}
\caption{Temperature dependence of the correlation plateaux $q_l$, 
for three types of correlations with $l=1,2,3$,
 computed using the cavity method. 
The main graph shows the continuous transition of
 strongly correlated HP sequences. 
The inset displays 
the discontinuous transition of anticorrelated HP sequences.}
\label{ord_para}
\end{figure}

The freezing phenomenon can be quantitatively described by the correlation plateaux
$q_l$ defined above. In Fig. \ref{ord_para} we present our analytical
results for $q_l(T)$ for two sequences with opposite correlations.

What are the modifications of the above mean field picture for 
realistic (finite-dimensional) models? 
At least two features must be seriously reconsidered. First, 
within our approximation, the formation of ordered structures 
is generally inhibited. Let us consider, for instance, the HP 
copolymer with highly correlated sequences $\pi\approx 1$. In dimension 
$D$, at low temperature, the system can separate into A-rich and 
B-rich regions of linear size $\ell\sim \log(1-\pi)^{1/D}$, 
with a frustration energy of order $\ell^{-1}$ per monomer. 
The phase diagram in Fig. \ref{transitions} must therefore  be 
modified in the $\pi\approx 1$ (for the HP model) and $\pi\approx 0$ 
(for the ampholyte) regions due to the emergence of spatially ordered phases.
A proper treatment of these phenomena (which are crucial for block 
copolymers~\cite{SfatosShakhnovich97})
is beyond the scope of this paper.

Second, the discontinuous dynamical phase transition to the frozen 
glass phase cannot survive in finite dimensions because of nucleation 
phenomena. According to the usual scenario
of discontinuous glass transitions, we
 expect it to become a sharp crossover associated 
with the emergence of very large (`activated') relaxation times~\cite{dyn_rev}.

In conclusion, we have described a new approach to 
heteropolymeric systems. 
The mean-field approximation is introduced in a well-defined and consistent 
way, and any further calculation can be checked through simulations on 
the Bethe lattice. The flexibility of the approach should allow for many applications (e.g., designed sequences) 
and further refinements (e.g., clusters of sites in the cavity).
We considered the case of copolymers with correlated
sequences and found that the frozen glass phase 
can be preceded by a soft glass phase with
very different dynamical properties. 
Since the sequence correlations turned out to be the decisive parameter, 
it would be interesting to analyze protein 
sequences in the light of our findings~\cite{ProteinCorrelation}.

\vspace{-0.2cm}
\bibliographystyle{prsty}

\end{document}